\documentclass[10pt,journal,compsoc]{IEEEtran}

\ifCLASSOPTIONcompsoc
  \usepackage[nocompress]{cite}
\else
  \usepackage{cite}
\fi
\usepackage{amsmath}
\usepackage{mathbbol}
\usepackage{booktabs}
\usepackage{soul}
\usepackage{url}
\usepackage{scalerel}
\usepackage{balance}
\usepackage{enumitem}
\usepackage{subfigure}
\usepackage[linesnumbered,ruled,vlined]{algorithm2e}
\usepackage{algpseudocode}
\usepackage{hyperref}
\usepackage{cleveref}
\usepackage{xcolor}
\usepackage[switch]{lineno}
\usepackage[utf8]{inputenc}

\usepackage{array}
\usepackage{multirow}

\crefformat{section}{\S#2#1#3} 
\crefformat{subsection}{\S#2#1#3}
\crefformat{subsubsection}{\S#2#1#3}

\begin{document}

\title{Training {Large-Scale} News Recommenders with Pretrained Language Models in the Loop}


\author{Shitao~Xiao,~Zheng~Liu,~Yingxia~Shao,~\textit{Member,~IEEE,}~Tao~Di,~Xing~Xie,~\textit{Member,~IEEE}~
\IEEEcompsocitemizethanks{
\IEEEcompsocthanksitem Shitao Xiao and Yingxia Shao are with BUPT\protect\\
E-mail: \{stxiao,shaoyx\}@bupt.edu.cn.
\IEEEcompsocthanksitem Zheng Liu and Xing Xie are with Microsoft Research Asia\protect\\
E-mail: \{zhengliu,xingx\}@microsoft.com
\IEEEcompsocthanksitem Tao Di is with Microsoft\protect\\
E-mail: Tao.Di@microsoft.com}
}

\markboth{Journal of \LaTeX\ Class Files,~Vol.~14, No.~8, August~2015}%
{Shell \MakeLowercase{\textit{et al.}}: Bare Demo of IEEEtran.cls for Computer Society Journals}

\IEEEtitleabstractindextext{%
\begin{abstract}
News recommendation calls for deep insights of news articles' underlying semantics. Therefore, pretrained language models (PLMs), like BERT and RoBERTa, may substantially contribute to the recommendation quality. However, it's extremely challenging to have news recommenders trained together with such big models: the learning of news recommenders requires intensive news encoding operations, whose cost is 
prohibitive if PLMs are used as the news encoder. In this paper, we propose a novel framework, {SpeedyFeed}, which efficiently trains PLMs-based news recommenders of superior quality. SpeedyFeed is highlighted for its light-weighted encoding pipeline, which gives rise to three major advantages. 
Firstly, it makes the intermedia results fully reusable for the training workflow, which removes most of the repetitive but redundant encoding operations. 
Secondly, it improves the data efficiency of the training workflow, where non-informative data can be eliminated from encoding. 
Thirdly, it further saves the cost by leveraging simplified news encoding and compact news representation.
Extensive experiments show that SpeedyFeed leads to more than 100$\times$ acceleration of the training process, which enables big models to be trained efficiently and effectively over massive user data. The well-trained PLMs-based model from SpeedyFeed demonstrates highly competitive performance, where it outperforms the state-of-the-art news recommenders with significant margins.
SpeedyFeed is also a model-agnostic framework, which is potentially applicable to a wide spectrum of content-based recommender systems; therefore, the whole framework is open-sourced to facilitate the progress in related areas. 

\end{abstract}

}

\maketitle

\IEEEraisesectionheading{\section{Introduction}\label{sec:introduction}}
Online news platforms have been important media of information acquisition. Given the huge volumes of online news articles, personalized news feed \cite{wu2020mind,okura2017embedding,li2010contextual} become imperative, with which users may get the news articles they feel interested in. The high-quality news recommendation is built upon the precise understanding of news articles' underlying semantics. Therefore, the pretrained language models (PLMs), e.g., BERT and RoBERTa \cite{devlin2018bert,liu2019roberta}, which achieve remarkable performances on general text understanding tasks, are desirable of being applied as the news encoder. However, the PLMs are not quite friendly to the end-to-end training of news recommenders. On the one hand, it is expensive to work with PLMs: the encoding speed will be relatively slow and the GPU RAM consumption will be huge given the considerable sizes of PLMs. On the other hand, the training of news recommenders requires intensive news encoding operations: to learn from every click signal of a user, the user's entire historical news clicks need to be encoded, whose computation cost will be prohibitive if PLMs are utilized. As a result, the development of PLMs-based news recommenders is severely limited by the efficiency bottleneck.

To overcome the above challenge, a novel framework SpeedyFeed is proposed in this work, which trains PMLs-based news recommenders with high efficiency and high quality. SpeedyFeed is highlighted for its light-weighted encoding pipeline, which leads to the following advantages. 

$\bullet$ {\em The intermedia results are made highly reusable}. Instead of having training instances encoded for one-shot use and discard afterwards, our framework saves the cost by making the following intermedia results fully reusable. Firstly, there are a small fraction of ``breaking news articles'', which are highly popular and widely exist in the majority of users' histories. Such news articles may frequently appear throughout the training process, and need to be re-encoded everytime. Knowing that the news recommenders are trained with small learning rates, usually in the magnitude of 1e$^{-5}$ (especially when PLMs are fine-tuned), the same news article's embedding will merely be slightly changed in every training step. As such, a caching mechanism is developed, which enables freshly generated news embeddings to be reused for multiple steps. Secondly, it is also wasteful of simply having user history encoded for the prediction of one single click signal. In our framework, the autoregressive user modeling is proposed, where an encoded prefix of user history can be reused for the calculation of all its subsequent user embeddings. 

$\bullet$ {\em The data efficiency is significantly improved.} The typical training workflow is prone to poor data efficiency: given that the lengths of news articles and user histories are highly diversified, plenty of padded elements have to be introduced so that raw data can be batched as input tensors. The padded data is not only non-informative, but also severely slows down the training speed. In our framework, a centralized news encoding workflow is designed, which completely eliminates the padded data in user history. Besides, the data loader is designed to adaptively group the training instances, so that less padded data is needed for the news articles. 

$\bullet$ {\em The news encoding cost is reduced with well preserved encoding quality.} The PLMs are limited by their quadratic encoding complexity, which makes the news encoding cost grow dramatically when the news article's length becomes longer. In our framework, two techniques are utilized to mitigate this problem. Firstly, the bus language modeling (BusLM) is introduced for news encoding: on the one hand, it partitions each news article into small segments, which results in the linear reduction of encoding complexity; on the other hand, it establishes the bus connection between the segments, which makes them jointly encoded for a high-quality news embedding. Secondly, the content refinement is performed for each news article before it is encoded by PLMs: the useful part of a news article is identified from the raw content, based on which the news article is transformed into a more compact representation.

It is worth noting that SpeedyFeed is not solely for training speedup. But because of the high training speed, it is now made feasible of training large-scale PLMs-based news recommenders over a huge amount of user data. The enlarged model scale, together with the enriched training data, ultimately make our recommender superior in generating high-quality news recommendation. SpeedyFeed is verified with the production data of Microsoft News, where it leads to more than 100$\times$ acceleration of the training speed, compared with its conventional workflow. Besides, our well trained PLMs-based recommender also demonstrates highly competitive performance, where it significantly outperforms the state-of-the-art approaches in comprehensive  evaluations. 

Finally, SpeedyFeed is a model-agnostic framework. In fact, it can be helpful to a wide variety of content-based recommender systems where user behaviors are associated with rich textual information, such as commodity and advertisement recommendation. 

We summarize the major contributions of this work with the following points:
\begin{itemize}
    \item We propose a novel framework SpeedyFeed to facilitate the training of large-scale PLMs-based news recommenders. With highly improved reusability, data efficiency and reduced news encoding complexity, our framework achieves significant acceleration for the training process.
    \item Our framework also fully preserves the PLMs' expressiveness. The PLMs-based news recommender, trained via SpeedyFeed, significantly outperforms the state-of-the-art news recommendation baselines with even smaller training cost.
    \item We've made SpeedyFeed\footnote{https://github.com/Microsoft/SpeedyRec} open to public, which can be adapted for a wide spectrum of content-based recommendation systems with rich textual information. 
\end{itemize}



\section{Related Works}
In this section, we briefly review the related works from two perspectives: the deep news recommendation systems, and the pretrained language models.

\subsection{Deep News Recommendation Systems}
News recommendation systems are designed to identify users' interested news articles with intensive exploitation of their historical news browsing behaviors \cite{wu2020mind,li2010contextual,das2007google}. As a result, two inherent problems need to be resolved within this process. One problem is the modeling of users' behaviors. With the progress of deep learning based recommendation systems, {plenty of} techniques have been proposed for user modeling. In Youtube-DNN \cite{covington2016deep}, users are represented as the averages of their interacted items' embeddings; in GRU4Rec \cite{hidasi2015session}, users' historical behaviors are aggregated with GRUs for sequential awareness; in DIN \cite{zhou2018deep} and DEIN \cite{zhou2019deep}, users' historical behaviors are attentively aggregated to establish candidate dependency; and in RUM~\cite{chen2018sequential}, memory networks are utilized to capture the diversity about users' behaviors. Such technical advancement also inspires the development of news recommenders. In DKN \cite{wang2018dkn}, users' historical news clicks are attended by the candidate news for more precise modeling of user interest; and in LSTUR \cite{an2019neural}, recurrent neural networks are leveraged to capture users' short-term interests.

The other problem, which is more specific to news recommendation, is the modeling of news content. In recent years, the prosperity of natural language processing pushes forward the progress of news modeling. For example, the hierarchical attention networks (HAN) \cite{yang2016hierarchical}, which was originally proposed for document classification, is adapted for the multi-view representation of news articles \cite{wu2019neural}; meanwhile, the Deep Attention Matching Networks (DAMN)~\cite{zhou2018multi}, which was designed for response selection in chatbots, is applied to perform fine-grained matching between the news content and user history. The remarkable progress of the pretrained language models brings huge potentials for the further enhancement of news modeling. However, the efficiency issue becomes one of the major obstacles of applying PLMs for news recommendation: compared with the conventional small-scale text encoders, making use of PLMs in news recommenders is expensive in terms of both training time and computation resources. It usually requires the models to be trained on powerful GPU clusters, and still takes tremendously more running time. As a result, the research progress and real-world applications of PLMs-based news recommenders are comparatively limited for the current stage.

\subsection{Pretrained Language Models}
The pretrained language models are proposed to learn universal representation/generation models with neural networks trained on large-scale corpus. The early works were started with some shallow structures, e.g, Skip-Gram \cite{mikolov2013distributed} and GloVe \cite{pennington2014glove}; in recent years, the network structures are being quickly scaled up: from EMLo \cite{peters2018deep}, GPT \cite{radford2018improving}, to BERT \cite{liu2019roberta}, RoBERTa \cite{liu2019roberta}, UniLM \cite{bao2020unilmv2}, till today's GPT-3 \cite{brown2020language}. The large-scale models, which get fully trained with massive corpus, demonstrate superior capabilities on general NLP tasks, e.g., semantic matching, question-answering, machine translation, and response generation.

\begin{figure*}[t]
\centering
\includegraphics[width=1.0\textwidth]{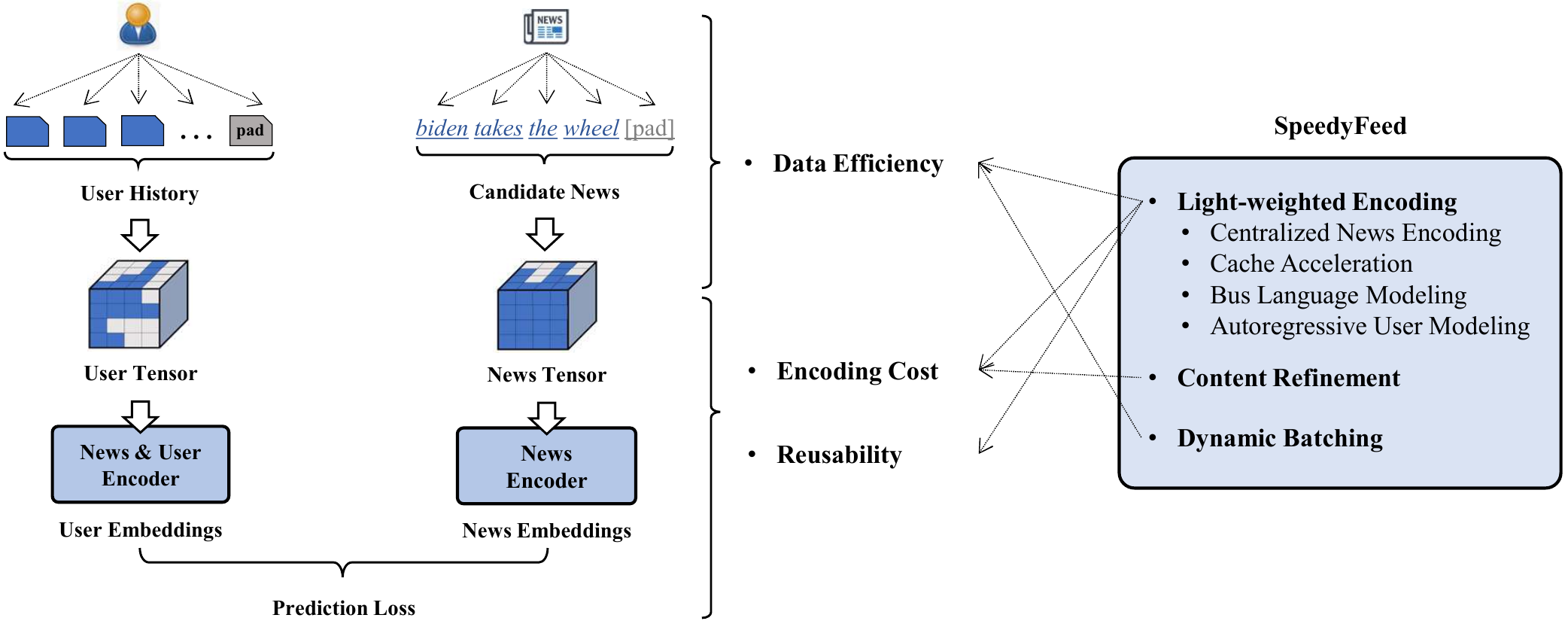}
\caption{Left: the typical training workflow of news recommendation. Middle: underlying problems within the typical workflow. Right: the constitution of SpeedyFeed, and how it contributes to the training workflow.}
\label{fig:news-reco}
\end{figure*}

The pretrained language models are also being intensively applied for the retrieval or information-filtering related scenarios \cite{chang2020pre,humeau2019poly,khattab2020colbert}; e.g., in \cite{guu2020realm}, PLMs are trained for knowledge retrieval, and in \cite{lu2020twinbert}, PLMs are fine-tuned for advertisement keyword matching. In these scenarios, PLMs are required to represent a query and a keyword into their latent embeddings, where the query-keyword relationship can be reflected by their embedding similarity. Apparently, news recommenders turn out to be similar applications. However, the PLMs-based news recommenders can be relatively more expensive: to match a user towards a candidate news, it needs to encode all of the user's historical news clicks with PLMs, which will lead to huge encoding costs.

\section{Preliminaries}
\subsection{Typical Workflow of Training News Recommenders}\label{sec:typical}
The news recommender is to predict user's future news preference given their news clicks in history. Therefore, a typical training workflow consists of three steps (shown on the left side of Figure \ref{fig:news-reco}), as implemented by Microsoft Recommenders \cite{msreco}.

 \textbf{1) Input Processing}. The trainer needs to transfer the raw data, i.e., user's historical interactions with news, into the required format, such that it can be loaded for training. Two operations are involved in this stage. On the one hand, the news articles are tokenized, and then padded/truncated into token sequences of a unified length. On the other hand, all users' histories are padded/truncated into news sequences of a unified length. 

 \textbf{2) News \& User Encoding}. The input tensors are encoded via two steps \cite{wu2020mind}. Firstly, the news encoding, which maps user's historical news clicks and all of the candidate news into news embeddings. Secondly, the user encoding, which generates user embeddings based on the encoded historical news and other auxiliary features. 

 \textbf{3) Learning from Prediction}. Finally, the prediction is made about which candidate news is clicked given the user and news embeddings. The prediction loss (e.g., cross entropy or BPR) will be back-propagated for the model parameters' update.

\subsection{What's wrong with the typical Workflow} One of the most notable things about training a news recommender is its huge text encoding cost: to make a prediction for one single user click, the trainer needs to encode 1) \textit{all of the news articles in user history} and 2) \textit{the candidate news}. Considering the large magnitudes of pretrained language models, the text encoding related computations will almost dominate the entire training cost. However, because of the following issues (shown in the middle of Figure \ref{fig:news-reco}), the typical training workflow becomes severely limited in efficiency, which makes it prohibitive to train a PLMs-based large-scale news recommender.

$\bullet$ \textbf{High Encoding Cost}. First of all, the PLMs are considerably larger in scale than the text encoders used in conventional text-related recommendations, e.g., bi-LSTM, CNN, and shallow transformers. What is worse, the PLMs are highly unfavorable to the processing of long texts. Particularly, the encoding cost is vulnerable to the length of input news ($N$), whose time complexity is $O(N^2)$, given that the mainstream PLMs are all based on transformer architectures \cite{vaswani2017attention}. Considering that many news articles require long textual descriptions to fully express their underlying information, it results in a huge computation overhead while encoding such news articles.

$\bullet$ \textbf{Low Reusability}. Secondly, the reusability is seldom emphasized before: every time a training instance is given, it is processed for the calculation of its own loss; once the loss is back-propagated, all related intermediate results, especially the news embeddings, will all be discarded after being used just one time. Considering that it is quite an expensive operation to encode a news article with PLMs, such a defect severely slows down the training progress.

$\bullet$ \textbf{Low Data Efficiency}. Lastly, due to the existence of substantial padded data (the padded elements in news article and user history), the meaningful computation throughput can be severely limited. Particularly, given that a token is the atomic data unit for both user tensor and candidate tensor, we define data efficiency as the ratio of valid (i.e., non-padded) tokens within the input tensors:
\begin{equation}\label{eq:de}
  \mathrm{DE} = \frac{|\text{valid tokens}|}
  {|\text{valid tokens}|+|\text{padded tokens}|} \times 100\%.
\end{equation}
Due to the highly diversified lengths of user histories and news articles, a huge amount of padded elements will probably be introduced. We empirically find that the data efficiency is usually lower than 50\% in practice, which leads to a big waste of computation capacity and further jeopardizes the training efficiency.

\section{Methodology}
We develop an efficient training framework SpeedyFeed, which enables news recommenders built upon large-scale PLMs to be trained with both high speed and high quality. With SpeedyFeed, the news and user encoding are carried out through a light-weighted encoding pipeline, which is characterized by the following techniques: 1) the centralized news encoding for high data efficiency, 2) the cache acceleration and the autoregressive user modeling for high reusability, 3) the bus language modeling for economic encoding complexity (as the rightmost of Figure \ref{fig:news-reco}). Besides, two auxiliary techniques: content refinement and dynamic batching, are introduced, which give rise to a more compact representation of news content and further reduction of padded data, respectively.

\subsection{Light-weighted Encoding Pipeline}

\begin{algorithm}[t]
\caption{Light-weighted Encoding Pipeline}\label{alg:1}
    \LinesNumbered 
    \KwIn{a mini-batch: user tensor \textbf{U}, news tensor \textbf{N}}
    \KwOut{Overall prediction loss $\mathcal{L}_{auto}$}
    \Begin{
        Merged set \textbf{M}:  gather(\textbf{U}.news $\cup$ \textbf{N}.\text{news})\;
        Cached set $\textbf{M}_{C}$: $\{ \text{\textit{m}: \textit{m} in cache} \}_{\textbf{M}}$\;
        Get lookup rate $p_t$ from scheduler (Eq. \ref{eq:schedule})\;
        Lookup Set $\textbf{M}_L$: sample from $\textbf{M}_C$ with $p_t$\;
        News embeddings set $\mathbf{\Theta}_1$
        $\leftarrow$
        CacheLookup($\textbf{M}_L$)\;
        News embeddings set $\mathbf{\Theta}_2$
        $\leftarrow$
        BusLM($\textbf{M} \setminus \textbf{M}_L$)\;
        Dispatch $\mathbf{\Theta}_1 \cup \mathbf{\Theta}_2$\;
        Refresh cache with $\mathbf{\Theta}_2$\;
        Get $\mathcal{L}_{auto}$ from autoregressive user modeling. 
    }
\end{algorithm}

{Algorithm \ref{alg:1} presents the main logics of the light-weighted encoding pipeline. 
For each mini-batch, we apply {\em centralized news encoding} (Section~\ref{sec:centralized_news}) to gather all the involved news articles from both user tensor and news tensor into the merged set \textbf{M}.
Then the lookup set $\textbf{M}_L$ is sampled from the cached news embeddings $\textbf{M}_C$ with the lookup rate $p_t$, and the news articles within the lookup set directly use their cached embeddings, denoted by $\boldsymbol{\Theta}_1$. The news articles out of the lookup set are encoded with BusLM (Section~\ref{sec:bus}), which gives $\boldsymbol{\Theta}_2$. The whole news embeddings $\boldsymbol{\Theta_1} \cup \boldsymbol{\Theta_2}$ are dispatched to their original positions (in either user history or candidate news); then, the cache is refreshed with the newly generated news embeddings $\boldsymbol{\Theta}_2$. Finally, the overall prediction loss is calculated with the autoregressive user modeling (Section~\ref{sec:auto}), as in Eq. \ref{eq:auto}.}

{In the following sections, we elobrate the details of technical contributions in the light-weighted encoding pipeline.}

\subsubsection{Centralized News Encoding}\label{sec:centralized_news}
The overall news encoding workflow is discussed in the first place. In the typical training workflow, the news encoder will directly work on the input tensors (i.e., user tensor, news tensor) for the news embeddings. During this process, the padded news articles are encoded together with the valid news, which results in low data efficiency.

\begin{figure}[t]
\centering
\includegraphics[width=0.83\columnwidth]{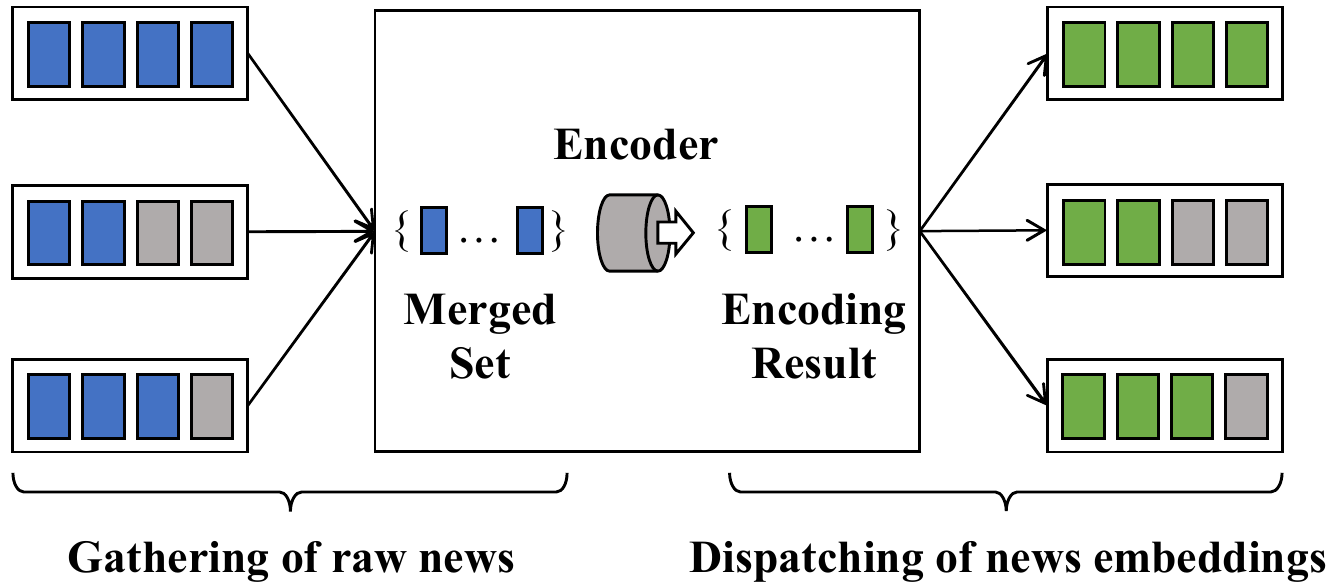}
\caption{Illustration of centralized news encoding: the news articles are gathered from the training instances, with all padding (grey) and duplicated ones removed from the merged set; the encoding results, i.e., the news embeddings (green), are dispatched to the training instances.}
\label{fig:encode}
\end{figure}

Unlike the typical method, all of the news articles within a common mini-batch are jointly encoded in SpeedyFeed (as Figure \ref{fig:encode}). The centralized news encoding takes 3 steps: gathering, encoding and dispatching. Once a mini-batch is given, it gathers the news articles from all users and candidates into a merged set. The padded news and duplicated news are all removed. Then, the new embeddings are generated for all remaining news in the merged set. Finally, the news embeddings are dispatched to their original training instances. Note that the padded news articles also require their embeddings so as to infer the user embeddings; in this place, a dummy vector is plugged into the positions taken by the padded news, whereby no additional encoding cost is needed.

\renewcommand{\arraystretch}{1.25}
\newcommand\ChangeRT[1]{\noalign{\hrule height #1}}
\subsubsection{{Cache-accelerated News Encoding}}\label{sec:cache}

\begin{table}[!h]
    \caption{The long-tail property about news click. The Top-$1\%$ popular news yields almost 60\% of the overall clicks.}
    \centering
    \begin{tabular}{c|c|c|c|c|c|c}
    \ChangeRT{1pt}
        Top-$\alpha$ & 1$\%$ & 3$\%$ & 5$\%$ & 10$\%$ & 20$\%$ & 30$\%$ \\
    \hline
        Click Ratio ($\%$) & 59.53 & 77.65 & 84.82 & 92.47 & 97.36 & 98.97
        \\

    \ChangeRT{1pt}
    \end{tabular}

    \label{tab:news-long}
\end{table}

\begin{figure}[h]
\centering
\includegraphics[width=0.81\columnwidth]{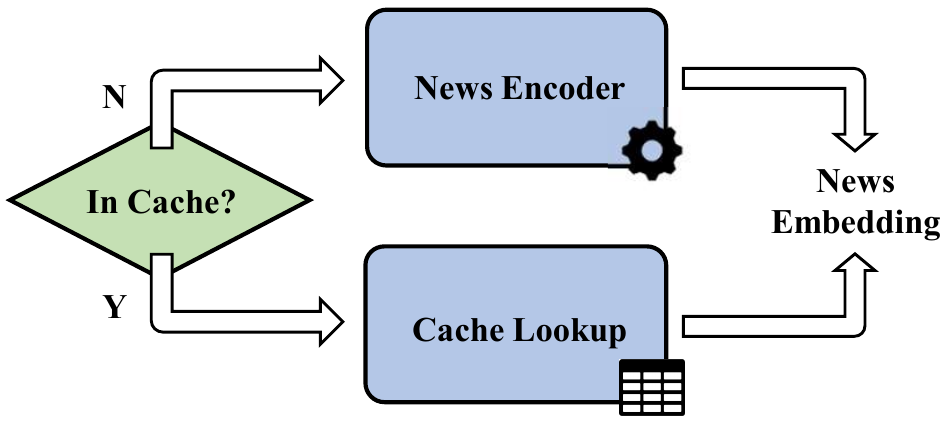}
\caption{Illustration of Cache-accelerated News Encoding: the in-cache news article's embedding is obtained by looking up the cache; the non-cached news article's embedding needs to be encoded from scratch.}
\label{fig:cache}
\end{figure}

The cache mechanism is developed to fully reuse the intermedia news encoding results. Particularly, one notable observation about Microsoft News is its long-tail property of the news click distribution. As shown in Table \ref{tab:news-long}, the top-$1\%$ popular news articles yield almost $60\%$ of the overall news clicks. Therefore, such popular news articles may widely exist in the majority of users' histories, making them frequently re-encoded across different training batches. Knowing that the model parameters are updated with a fairly small learning rate, usually in the magnitude of $1e^{-5}$, one news article's recent embedding can be reused in the current mini-batch for approximation. Based on this intuition, we propose Cache-accelerated News Encoding, where a cache is maintained in memory for the storage of fresh news embeddings. The news encoding workflow is changed accordingly as Figure \ref{fig:cache}.

$\bullet$ \textbf{News Encoding with Cache}. For each news article in a mini-batch, the trainer will check the cache in the first place: if there is a copy of news embedding in cache, it will be directly reused without encoding; otherwise, the news article will be encoded from scratch.

\begin{figure}[!h]
\centering
\includegraphics[width=0.75\columnwidth]{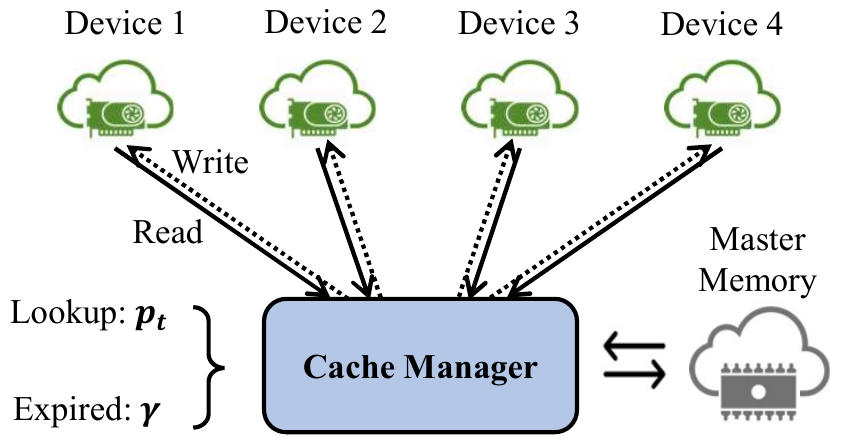}
\caption{Illustration of Cache Management: in the initial stage, the cached news embeddings are looked up with probability $p_t$; the cached news embeddings will be expired after $\gamma$ steps; all devices will share a common cache maintained in the master node's memory.}
\label{fig:cache-manage}
\end{figure}

$\bullet$ \textbf{Cache Management Policy}. The cache is managed with the following principles. Firstly, all of the embeddings in cache must be newly generated in the past few steps; otherwise, it will be incompatible with the current model. Secondly, the cache lookup should be dynamically scheduled: in the initial stage, the cached news embeddings should be used with a relatively low probability, because the step-wise change of model parameters is sharp; as the training goes on, the lookup rate should be gradually increased, because the change of model parameters becomes mild. 

Based on the above principles, the cache management policy is made (as Figure \ref{fig:cache-manage}), which is subject to two decisive variables: the stepwise lookup rate $p_t$, and the expiration step $\gamma$. 1) An exponential scheduler is used to control the probability of whether to lookup the cache: the cache is looked up with probability 0 when the training gets started; the lookup probability will gradually grow to $p_t$ at the $t$-th step w.r.t. the following relationship:
\begin{equation}\label{eq:schedule}
    p_t = 1.0 - \exp(-\beta t).
\end{equation} 
$\beta$ is the hyper parameter for growth rate, which lets $ p_t$ grow to 1.0 after the initial stage of the training process. 
2) A cached news embedding is expired after $\gamma$ steps, and then it is removed from the cache. 

Finally, instead of maintaining a private cache for each training thread, we establish a global cache in the master node. As a result, the newly generated news embeddings in one node can be shared across all devices, which accommodates the distributed training of news recommender. Besides, the cache is maintained in memory, instead of GPU RAM; therefore, it is almost free of cost, and the storage capacity can be large enough to host all needed embeddings.

\textbf{{Summary}}.
{We summarize the cache mechanism in Algorithm~\ref{alg:cache}. In the training step $t$, we input a merged news set $\mathbf{M}$ by centralized news encoding. Firstly, the lookup probability $p_t$ is generated according to the current step $t$ and hyper-parameter $\beta$ (Lines 3). We use a random value to determine whether to read embeddings from the cache. If true, for all news embeddings in the cache, we only load the ones which are encoded less than $\gamma$ training steps before the current step (Line 8). These loaded embeddings are denoted by $\mathbf{\Theta}_1$, and the corresponding news are denoted by $\mathbf{M}_L$. For news which are not cached, i.e., $\mathbf{M} \setminus \mathbf{M}_L$, we encode them into embeddings $\mathbf{\Theta}_2$ with BusLM, which is introduced in the next subsetction, and write $\mathbf{\Theta}_2$ into cache. Finally,  $\mathbf{\Theta}_1\cup \mathbf{\Theta}_2$ is the whole new embeddings in step $t$.}

\begin{algorithm}[t]
\caption{Cache-accelerated News Encoding}\label{alg:cache}
    \LinesNumbered 
    \KwIn{The merged news set $\mathbf{M}$ in mini-batch, current training step $t$}
    \KwOut{News embeddings $\mathbf{\Theta}$ for all news in $\mathbf{M}$}
    \Begin{
        initialize  $\mathbf{\Theta}_1$ and $\mathbf{M}_L$;\\ 
        $p_t$ = $1-\exp(-\beta t)$;\\
        \If{$random<p_t$}{
        \For{$n$ in $\mathbf{M}$} {
        \If{$n$ in cache} {
            $e_n$ $\gets$ Read(cache, n);\\
            \If{$t-t_{e_n}\leq \gamma$} 
            {   $\mathbf{\Theta}_1$.add($e_n$);\\ $\mathbf{M}_L$.add($n$);\\
            }
        }}}
        $\mathbf{\Theta}_2$ $\gets$ BusLM($\mathbf{M\setminus \mathbf{M}_L}$); \\
        Write(cache,  $\mathbf{\Theta}_2$); \\
        $\mathbf{\Theta}=\mathbf{\Theta}_1\cup \mathbf{\Theta}_2$.\\
    }
\end{algorithm}

\begin{figure*}[t]
\centering
\includegraphics[width=0.67\textwidth]{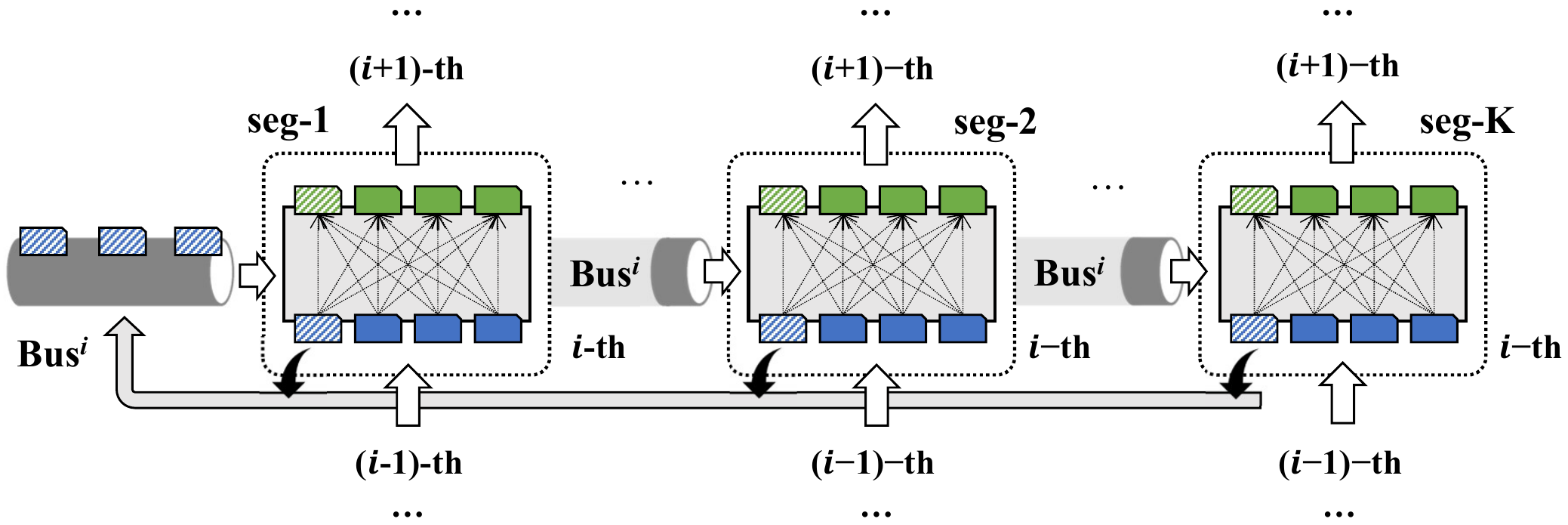}
\caption{Illustration of BusLM (using the $i$-th layer for demonstration): the first tokens from all segments are gathered as the bus of the $i$-th layer, $\text{Bus}^i$; $\text{Bus}^i$ is broadcasted to all segments, such that for each segment, the transformer attends to both the in-segment elements and the bus elements.}
\label{fig:bus}
\end{figure*}

\subsubsection{Bus Language Modeling}\label{sec:bus}
We make further analysis of how we conduct news encoding in an economic way. The news encoding complexity is to the square of news length $O(N^2)$. A straightforward way of time reduction is to split the news into several sub-components, e.g., the title, abstract, and body, as done in \cite{wu2019neural}; the text segments are processed independently, whose encoding results will be aggregated for the final news embedding. The operation may cut down the time complexity to $O(N^2/K)$, if the text can be partitioned into $K$ ``almost equal-length'' segments. Yet, the naive split of text harms the news embedding quality, as the text segments cannot make reference to each other during the encoding process. 

Inspired by recent progress on efficient transformers \cite{tay2020efficient}, we propose BusLM (Figure \ref{fig:bus}) to encode the news articles, where the acceleration is achieved with fully preserved embedding quality. In BusLM, the input news is uniformly partitioned into $K$ text segments, such that the encoding complexity is reduced to $O(N^2/K)$. The segments are still encoded by transformers; however, a layer-wise ``bus connection'' is established between the transformers, which enables information to be exchanged across the segments.

In each layer of the transformers, a ``proxy embedding'' is chosen for each segment, which serves as the sketch of its underlying information. To avoid additional computation as much as possible, we directly select the first embedding of each segment as its proxy; e.g., for the $i$-th layer of segment $j$, $\mathbf{H}^i_j[0]$ is chosen as the proxy (let $\mathbf{H}^i_j$ be the $j$-th segment's embedding sequence on the $i$-th layer). The $i$-th layer's proxy embeddings from all of the segments are gathered as the $i$-th Bus:
\begin{equation}
    \mathbf{Bus}^i = \{\mathbf{H}^i_j[0]\}^K_{j=1}.
\end{equation} 
The bus is used as a medium of information exchange, which all the segments may attend to directly. Particularly, the $i$-th to $i+1$-th transformation of the $j$-th segment will become:
\begin{equation}\label{eq:bus}
    \mathbf{H}^{i+1}_j = \mathrm{Transformer}^{i}(\big[\mathbf{H}^i_j, ~\mathbf{Bus}^i\big]),
\end{equation}
in which ``$[]$'' denotes concatenation, and ``$\mathrm{Transformer}^{i}(\cdot)$'' is the $i$-th layer of the transformers. The final news embedding is acquired by aggregating all of the hidden states in the last layer, i.e., $\textbf{H}_{*}^{-1}$. 
{In Appendix~\ref{apx:blm}, we present the implementation details of applying BusLM to different types of layers in transformer.}
It is empirically verified that both time-efficiency and memory-efficiency benefit from BusLM; meanwhile, the information loss due to the split of text is fully mitigated with the Bus connection. 

\begin{figure}[!t]
\centering
\includegraphics[width=1.0\columnwidth]{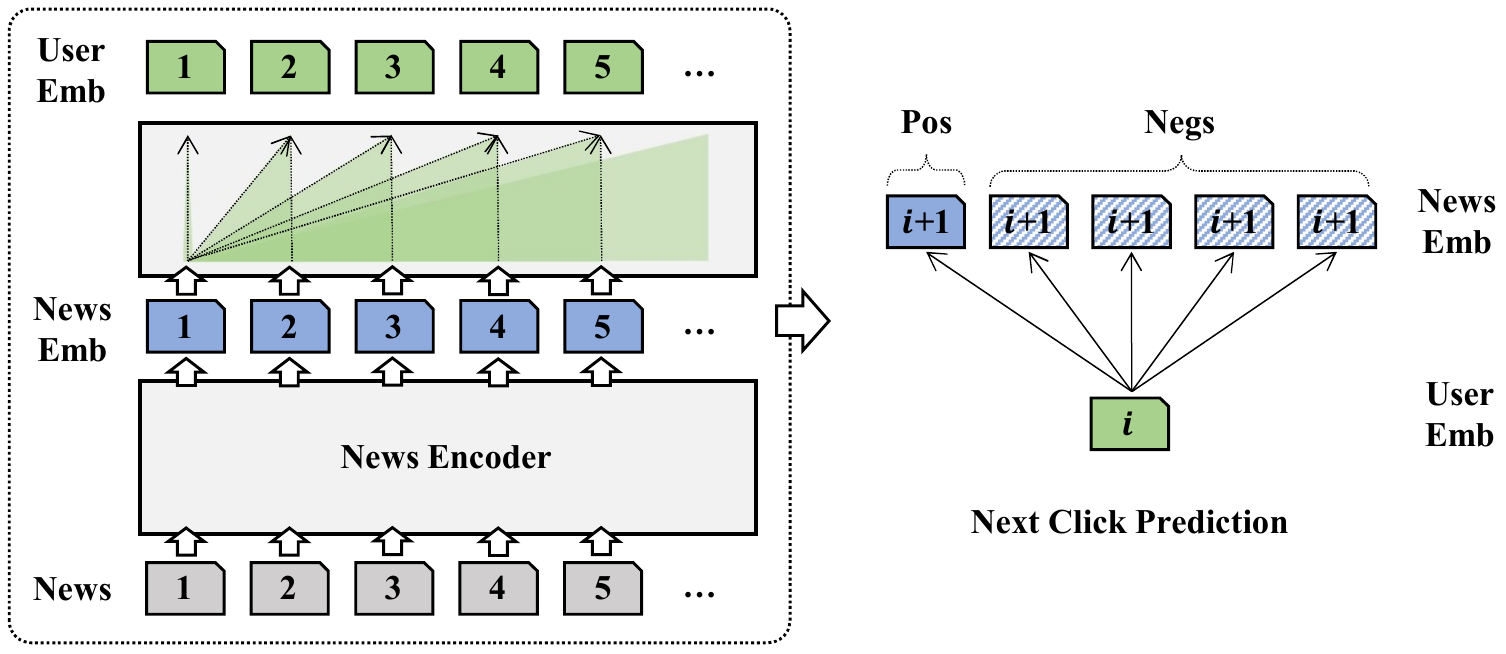}
\caption{Illustration of Autoregressive User Modeling: the user embeddings are generated for all timestamps, each of which is based on its current and preceding news embeddings (Left); each user embedding is used to predict its next news click (Right).}
\label{fig:auto}
\end{figure}

\subsubsection{Autoregressive User Modeling}\label{sec:auto}
The computation cost is further saved by reusing the encoded prefix of user history. As discussed, the news recommender is trained based on the prediction loss of user's news click. Therefore, a typical training instance consists of user's news click at one timestamp $t$ and its preceding news clicks: $\langle \mathrm{click}_{=t}, \mathrm{clicks}_{\leq t-1} \rangle$. However, when another training instance $\langle \mathrm{click}_{=t+1}, \mathrm{clicks}_{\leq t} \rangle$ is presented, $\mathrm{clicks}_{\leq t-1}$ becomes the prefix of user history, which requires to be re-encoded.

We propose autoregressive user modeling for more efficient utilization of user history (Figure \ref{fig:auto}), where all of the news clicks about a user can be predicted at one-shot of news encoding. Instead of processing each training instance $\langle \mathrm{clicks}_{=t}, \mathrm{click}_{\leq t-1} \rangle$ case-by-case, the whole user history $\mathrm{clicks}_{\leq L}$ will be treated as one unified training instance (let $L$ be the max length of user history). The trainer will encode all of the historical news clicks, which gives the news embedding set: $\{\boldsymbol{\theta}_l\}_{\leq L}$. The trainer will then calculate the user embedding set $\{\boldsymbol{\mu}_t\}_{\leq L}$, where $\boldsymbol{\mu}_t$ is conditioned on the preceding news embeddings $\{\boldsymbol{\theta}_l\}_{\leq t}$. Each user embedding is used to predict the news click for the next timestamp, where the overall prediction loss $\mathcal{L}_{auto}$ w.r.t. one sample user is computed as:
\begin{equation}\label{eq:auto}
    \mathcal{L}_{auto} = - \sum_{t < L} \log
    \frac{\exp(\langle \boldsymbol{\theta}_{t+1},\boldsymbol{\mu}_t \rangle)}
    {\exp(\langle \boldsymbol{\theta}_{t+1},\boldsymbol{\mu}_t \rangle) + 
    \sum_{ \boldsymbol{\theta}^{'}_{t+1} } \exp(\langle \boldsymbol{\theta}^{'}_{t+1},\boldsymbol{\mu}_t \rangle) }.
\end{equation}
``$\langle \cdot \rangle$'' calculates the relevance of the user and news embeddings, e.g., inner product; and $\boldsymbol{\theta}^{'}_{t+1}$ is the embedding of a negative sample.

\subsection{Further Enhancement}

\begin{figure}[!h]
\centering
\includegraphics[width=0.95\columnwidth]{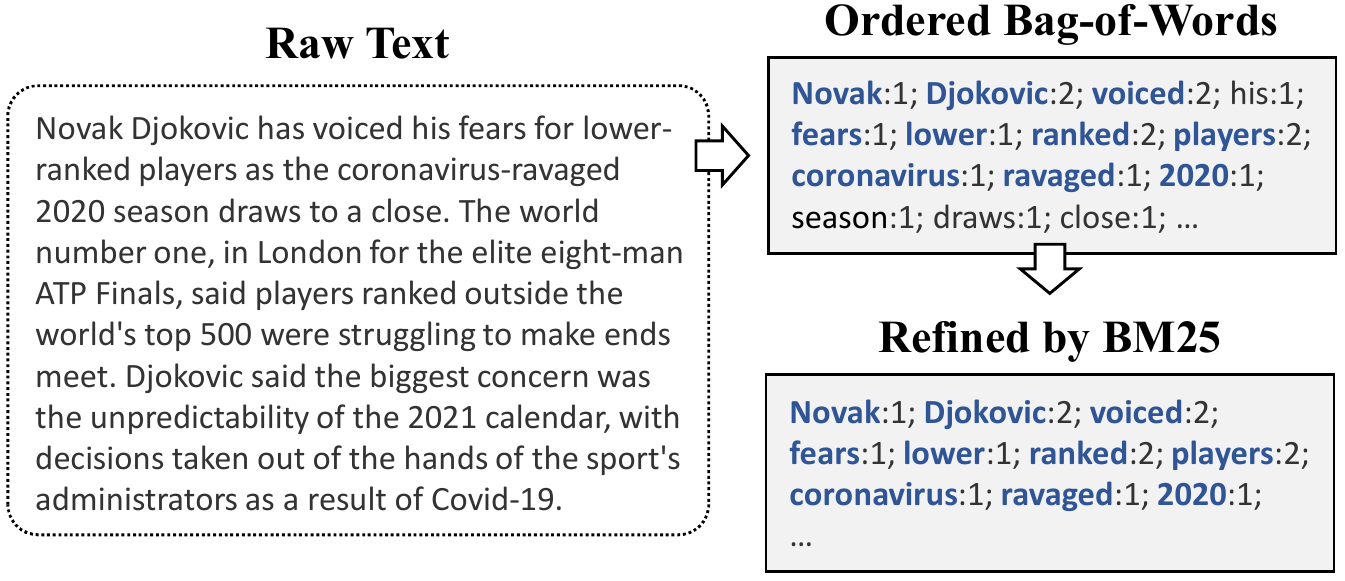}
\caption{Illustration of Content Refinement: the raw text is represented as an ``ordered-bag-of-words'' in the first place, with all stopping words and special characters removed; then, the top-$k$ important words are preserved based on their BM25 scores.}
\label{fig:refine}
\end{figure}

\subsubsection{Content Refinement}
Although the news encoding cost is reduced with BusLM, it is still prohibitive to process extremely long news articles. Therefore, moderate truncation of the new article is inevitable in practice. Instead of simply taking the head of a news article, the truncation is treated as a filtering operation in SpeedyFeed: we will try to remove the redundant or non-informative data, and preserve the important information as much as possible. The Ordered Bag-of-Words (OBoW) model is proposed for this purpose (Figure \ref{fig:refine}), which compactly represents the informative data distilled from the raw content. In OBoW, all special characters and stopping words are discarded; besides, the news article is represented as a sequence of ``{word}: {count}'' tuples ordered by the words' first-appearances in the original content. The remaining words are labeled with their BM25 scores. The words with the top-$k$ BM25 importance are believed to be the most informative, which will be preserved by the final OBoW model. 

One minor modification is required to encode the OBoW with PLMs: apart from the original token embedding, position embedding, and segment embedding, an additional frequency embedding is added to each input token, which brings in the information about each word's times of appearance in the original content.


\subsubsection{Dynamic Batching}
Dynamic batching is an asynchronous data loading process, which runs 
in parallel to the model training process. It uses multiple threads to read the user log and generate training instances (user history and candidate news) from it. The training instances are gathered as mini-batches and consumed by the training process. To reduce the number of padded tokens and maximize the GPU utilization, the following treatments are adopted. 

Firstly, the training instances are grouped based on the lengths  of their included news articles. As each training instance is virtually a collection of news articles, they can be marked by the ``max-length'' of their included news; e.g., an instance with 3 news, whose lengths are (32, \underline{48}, 36), will be marked with 48. Each training instance is routed to a bucket based on its max-length; e.g., the above instance may be routed to the bucket which hosts training instances of max lengths 40$\sim$50. All the news within one bucket will be padded to the same length based on the currently longest news in the bucket. Secondly, the bucket is checked after every fill-in: whether the total number of tokens reaches the threshold determined by the GPU RAM's capacity. Once a bucket is full, all of its training instances will be dump into a mini-batch and appended to the mini-batch queue, which is consumed by the training process. 

The above generation of mini-batch is ``dynamic'', as the padded length and the batch size are determined case-by-case. Based on the grouping operation, the overall padded length can be minimized; and with the dynamic batch size, the GPU capacity can be used as much as possible.

\section{Experiments}

\subsection{Settings}

\subsubsection{Dataset description}
We use a large real-world dataset for evaluation. The dataset contains Microsoft News\footnote{https://microsoftnews.msn.com/} users' news reading behaviors from 2020-05-01 to 2020-08-31. The data in the first 3 months (2020-05-01 to 2020-07-31) is used for training; the last month's data (2020-08-*) is used for testing. There are 1,202,576 news articles and 4,720,192 users, and it yields 72,093,576 news clicks in total (summarized as Table \ref{tab:dataset}). The following features are utilized for offline experiments. The news articles consist of their titles, abstracts, and bodies; the users are represented with their historical news clicks. Other features, like news categories, user demography and contexts, are omitted here, but exploited in production. More specifications of the data related to reproducibility are included in Appendix~\ref{apd:dataset}. 

\renewcommand{\arraystretch}{1.5}
\begin{table}[t]
    \caption{Statistics of the experimental dataset}
    \centering
    \begin{tabular}
    {>{\centering}p{0.16\columnwidth}>{\centering}p{0.16\columnwidth}>{\centering}p{0.2\columnwidth}>{\centering\arraybackslash}p{0.2\columnwidth}}
    \ChangeRT{1pt}
        \#User & \#News & \#Impressions & \#Clicks \\
    \hline
        4,720,192 & 1,202,576  & 41,481,252 & 72,093,567 \\
    \ChangeRT{1pt}
    \end{tabular}

    \label{tab:dataset}
\end{table}

\newcolumntype{C}[1]{>{\centering\let\newline\\\arraybackslash\hspace{0pt}}m{#1}}
\begin{table*}[t]
\caption{Upper: baselines w.o. SpeedyFeed acceleration. Lower: PLMs-based news recommenders accelerated with SpeedyFeed.}
    \centering
    \begin{tabular}{C{1.8cm} | C{1.2cm} C{1.2cm} C{1.5cm} C{1.7cm}  | C{1.6cm} C{1.8cm} C{1.8cm} | C{1.8cm} }
    \ChangeRT{1pt}
        & \textbf{\small{AUC}} & \textbf{\small{MRR}} & \textbf{\small{NDCG@5}} & \textbf{\small{NDCG@10}} & \textbf{\small{Recall@50}} & \textbf{\small{Recall@100}} & \textbf{\small{Recall@200}} & \textbf{\small{Time}} \small{(hour)} \\
    \hline
       {NPA} & 65.01 & 24.66 & 26.06 & 30.90 & 2.24\% & 4.04\% & 7.14\% & 23.6 \\
       {NAML} & 67.57 & 26.90 & 28.73 & 33.75 & 2.29\% & 5.22\% & 8.67\% & 26.6\\
       {LSTUR} & 64.37 & 24.35 & 25.58 & 30.63 & 2.28\% & 3.96\% & 6.84\% & 30.4 \\
       {NRMS} & 68.62 & 27.30 & 29.09 & 34.15 & 3.10\% & 5.81\% & 9.15\% & 27.8 \\
       {UniLM} & -- & -- & -- & -- & -- & -- & -- & 2497.5 \\
    \hline
       Speed-Mini & 72.06 & 30.16 & 32.63 & 37.74 & 6.93\% & 10.75\% & 16.13\% & {3.1} \\
       Speed-Half & 69.47 & 28.06 & 30.04 & 35.11 & 4.78\% & 7.47\% & 11.84\% & 3.3 \\
       Speed-Last & 70.98 & 28.92 & 31.09 & 36.26 & 4.69\% & 8.07\% & 12.94\% & 6.5 \\
       Speed-UniLM & \textbf{73.74} & \textbf{31.70} & \textbf{34.40} & \textbf{39.53} & \textbf{8.32}\% & \textbf{13.17}\% & \textbf{19.40}\% & 19.4 \\
    \ChangeRT{1pt}
    \end{tabular}
    
    \label{tab:exp-main}
\end{table*}


\subsubsection{SpeedyFeed Recommenders}
A {default news recommender} is trained by SpeedyFeed, whose configurations are listed as follows.

$\bullet$ \textbf{News Encoder}. Our news encoder is initialized with the pretrained checkpoint of UniLMv2-base \cite{bao2020unilmv2} (\textbf{UniLM} for short), which is a 12-layer \& 768-hidden-dimension language model trained by Microsoft. Leveraging the state-of-the-art pretraining techniques, it outperforms other well-known PLMs of the same scale, e.g., BERT-base and RoBERTa-base, on GLUE benchmark and many other general NLP tasks. 

$\bullet$ \textbf{User Encoder}. A highly simplified user encoder is adopted by the default news recommender, as we target on the impact brought by the news encoder. Particularly, a simple adaptation of the YouTube-DNN \cite{covington2016deep}, namely, Attentive YouTube-DNN, is utilized: it makes use of the weighted-sum of user's historical news embeddings for user representation, where a learnable attention vector is introduced to generate the aggregation weights. 

Besides, we also combine the default user encoder with the following alternative news encoders, which are certain sorts of simplifications of the original UniLM.

$\bullet$ \textbf{MiniLM} \cite{wang2020minilm}, a high-quality distillation of UniLM; both its depth and width are reduced to 50\% of the original model (i.e., with 6 layers and 384 hidden-dimensions). 

$\bullet$ \textbf{UniLM-Half}, where the model scale is the same as MiniLM, but the model weights are directly inherited from the original UniLM. 

$\bullet$ \textbf{UniLM-Last}, which uses the whole UniLM but simply finetunes the last layer in the training stage. (It is different from the default one where UniLM is trained end-to-end.) Although it does not contribute to the feed-forward speed, it reduces the training cost as most of the layers are frozen: the GPU RAM usage will be lower, so that we may use larger batch sizes for acceleration. Besides, it also saves the cost 
as much fewer model parameters call for update. 

All the above approaches are trained with SpeedyFeed, thus referred {to} as Speed-UniLM, Speed-Mini, Speed-Half, and Speed-Last in the experiments.

\subsubsection{Baselines}
The following representative news recommender baselines are utilized in our experiments. 

$\bullet$ \textbf{NPA} \cite{wu2019npa}, which leverages personalized attention to select and aggregate useful information in user history.

$\bullet$ \textbf{NAML} \cite{wu2019neural}, which uses multi-view attention to aggregate user's historical news clicks.

$\bullet$ \textbf{LSTUR} \cite{an2019neural}, which relies on multiple neural network structures to jointly capture user's long-term and short-term interest.

$\bullet$ \textbf{NRMS} \cite{wu2019nrms}, which makes use of multi-head self attention to improve the quality of user representation.

The above approaches make trial of various user modeling strategies for news recommendation. But one thing in common is that all of them make use of comparatively small-scale text encoders to generate news embeddings, such as 1D-CNN or self-attention. These methods are trained following the default workflow as demonstrated in Microsoft Recommenders\cite{msreco}.


\subsubsection{Evaluations}
{The experiment results are comprehensively evaluated from three perspectives.}

1) \textbf{Ranking}: The evaluation is made for the ranking performance: given a testing impression, the recommender is required to generate the ranking orders for the impressed news articles; the ranking orders are compared with the ground-truth (i.e., the clicked news within the impression), whose performance is measured with the typical ranking metrics: \textbf{AUC}, \textbf{NDCG}, \textbf{MRR}. 

2) \textbf{Recall}: The evaluation is made for the recall performance: based on the user embedding generated by the recommender, the relevant news articles are retrieved from the whole production index. Since the relevance between user embedding and news embedding is measured by inner-product, it turns out to be a Max-Inner-Product-Search problem, where HNSW \cite{malkov2018efficient} is used as the backbone of ANN index. The performance is measured with \textbf{Recall@K}, where the ground-truth is still the clicked news of the testing impression. 

3) \textbf{Training Efficiency}: We also evaluate the {training efficiency}, where the time cost is measured with the following configurations.

\subsubsection{Training configurations}
All the training jobs are performed on an Azure\footnote{https://azure.microsoft.com/en-us/services/machine-learning/} machine, with 4$\times$Nvidia-V100-32G GPUs, 40$\times$Intel(R) Xeon(R) Platinum 8168 CPU @ 2.70GHz processors, run on Ubuntu 16.04.6. The models are implemented with PyTorch 1.7.0. More specifications about the training process are included in Appendix~\ref{apd:train}. 

\subsection{Experiment Analysis}

\subsubsection{Overall Performance}
The main experiments are performed to clarify the following issues: 1) the effect on recommendation quality when PLMs are utilized as the news encoders, and 2) the effect on efficiency when SpeedyFeed is leveraged for recommender training. The following conclusions can be drawn based on the experiments results reported in Table \ref{tab:exp-main}.

Firstly, our default recommender, which is equipped with a full-scale and end-to-end trained UniLM, beats all models with simplified (MiniLM, UniLM-half) or insufficiently trained (UniLM-last) UniLM. Besides, it outperforms all of the baseline recommenders by huge margins: all the ranking metrics are significantly improved, and the recall metrics go above the baselines by several times. Therefore, it validates {1) the recommendation quality can be greatly improved by large-scale PLMs}, and 2) {the PLMs need to be fully trained within the recommender so as to achieve the best performance}. 

Notice that the above comparisons do not deprecate the importance of user encoder. Instead, it points out the necessity of collaborative utilization of large-scale news encoders and expressive user encoders; otherwise, the recommendation quality will be severely limited due to the insufficient modeling of news semantic.

Secondly, our default recommender (which is larger than all baseline recommenders by several orders) can be efficiently trained with even less time compared with those small-scale recommenders trained by the conventional workflow. Besides, the recommenders with simplified UniLMs can be trained by SpeedyFeed with further less time, as the news encoding costs become relatively smaller. We also let the default recommender trained by the conventional workflow; and the training time becomes {2497.5} hours (estimated from a faction of training progress, i.e., \textit{average-time-per-step} $\times$ \textit{the-total-required-steps}). In other words, the training speed is accelerated by over {100$\times$} with SpeedyFeed. 


Finally, one additional issue is that whether we should resort to the distilled PLMs for further training speedup. Given the results in Table \ref{tab:exp-main}, we incline to stay with the full-scale PLMs (i.e., UniLM \textit{v.s.} MiniLM), as the recommendation quality can be significantly improved with an acceptable increment of training cost. Besides, although MiniLM is faster, the online inference speed is hardly a bottleneck for the recommendation of Microsoft News: there are merely tens of thousands of fresh news articles generated each day, which can be encoded and cached for the recommender with very little cost. However, we do not exclude distilled PLMs' necessity for other scenarios, like search and ads, where fresh contents are generated rapidly and need to be processed in realtime.  

\begin{table}[t]
    \caption{Left: default recommender's training time w./w.o. being accelerated by SpeedyFeed. Right: the overall speedup effect, and the speedup effect from each basic module.}
    \label{tab:exp-speed}
    \centering
    \begin{tabular}{p{4.0cm} p{3.2cm}  }
    \ChangeRT{1pt}
       \textbf{Training Time} & \textbf{Speedup} \\
       \hline
       w.o. SpeedyFeed: \underline{2497.5} hours & Overall Speedup: 128.7$\times$ \\
       w. SpeedyFeed: 19.4 hours & Central \& Batch: 3.0$\times$ \\
       & Cache: 1.98$\times$\\
       & Autoregressive: 17.0$\times$\\
       & BusLM: 1.27$\times$ \\
    \ChangeRT{1pt}
    \end{tabular}

\end{table}
\begin{figure}[t]
\centering
\includegraphics[width=0.98\columnwidth]{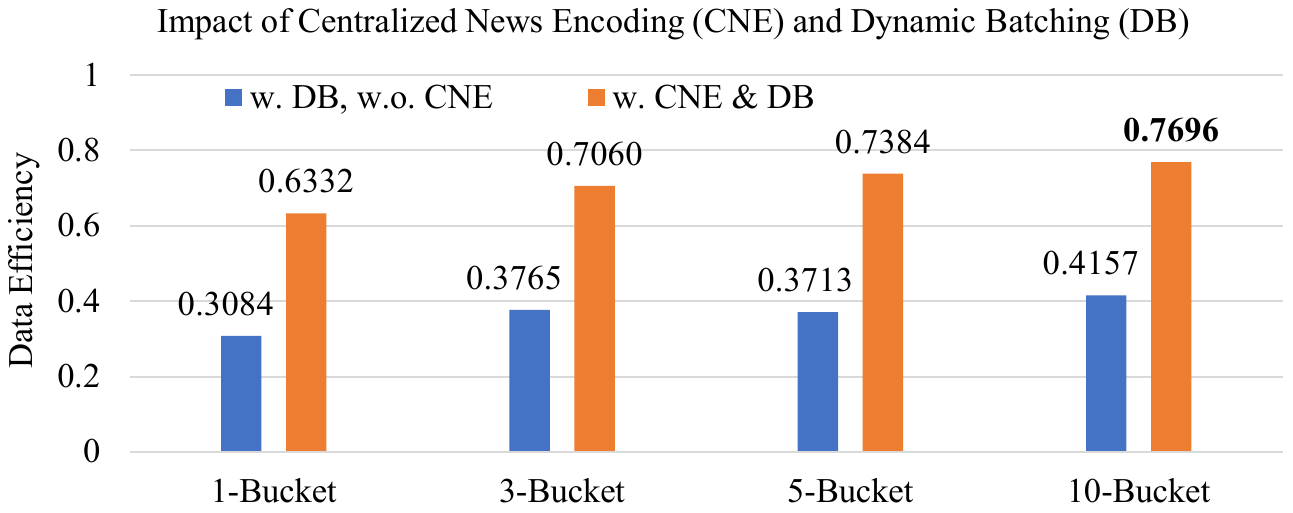}
\caption{Effect of Dynamic Batching (DB) and Centralized News Encoding (CNE) on data efficiency.}
\label{fig:exp-de}
\end{figure}

\begin{table}[t]
    \caption{Ablation studies, where BusLM, Cache-accelerated news encoding, and content refinement are disabled, respectively.}
    \label{tab:exp-abl}
    \centering
    \begin{tabular}{C{1.6cm} | C{1.2cm}C{1.6cm}C{1.8cm}  }
    \ChangeRT{1pt}
        & \textbf{\small{AUC}} & \textbf{\small{Recall@50}} & \textbf{\small{Recall@100}} \\
    \hline
       w.o. Bus  & 73.20 & 6.72 & 10.81 \\
       w.o. Cache  & 73.70 & {8.11} & {12.73} \\
       w.o. Refine & 73.70 & 7.88 & 12.45 \\
       Default     & \textbf{73.74} & \textbf{8.32} &       
                     \textbf{13.17} \\
    \ChangeRT{1pt}
    \end{tabular}

\end{table}

\subsubsection{Further Analysis} 
Experiments are performed to further clarify the following issues: 1) SpeedyFeed's impacts on training speedup, 2) SpeedyFeed's impact on data efficiency, 3) SpeedyFeed's impact on recommendation quality, 4) detailed analysis of cache-accelerated news encoding, and 5) detailed analysis of BusLM.

$\bullet$ \textbf{Impact on Speedup} We study the overall speedup effect of SpeedyFeed (Table \ref{tab:exp-speed}): the training time is reduced from 2497.5 hours (estimated) to 19.4 hours, which means a {128.7$\times$} speedup. We further analyze the detailed speedup effect of each module. We find that the autoregressive user modeling leads to the biggest gain, where the training speed is accelerated by 17$\times$. This observation is natural to interpret, as the encoding cost for the entire prefix of user history can now be saved by reusing the preceding encoding results (as discussed in Section \ref{sec:auto}). The centralized news encoding (Central) and the dynamic batching (Batch) jointly improve the data efficiency, which results in another 3$\times$ speedup. Besides, the cache-accelerated news encoding (Cache) and BusLM generate 2$\times$ and 1.27$\times$ speedup, respectively. 

$\bullet$ \textbf{Impact on Data Efficiency}. 
The data efficiency (Eq. \ref{eq:de}) is jointly increased with the centralized news encoding and the dynamic batching (Figure \ref{fig:exp-de}). The original data efficiency (1-Bucket, w.o. CNE) is merely around 30\%, which means roughly 70\% of the computation is wasted on the padded data. With the adoption of both techniques, a large portion of the padded data is removed, and the data efficiency can be easily improved to more than 70\%.

$\bullet$ \textbf{Impact on Recommendation Quality}. 
Shown as Table \ref{tab:exp-abl}, the recommendation quality is reduced when bus and content refinement are disabled. Particularly, for ``w.o. Bus'', the bus connection is removed, where the partitioned news segments become independently encoded. Without making reference to the whole context, it's hard to fully express each segment's underlying semantic, which harms the quality of news embedding. For ``w.o. refine'', the front part of each news article is truncated for input; therefore, it will result in more severely information loss, which reduces the recommendation quality.

\begin{table}[t]
    \caption{Effect of cache-accelerated news encoding. The cache is disabled when $\gamma=0$.}
    \label{tab:exp-cache}
    \centering
    \begin{tabular}{C{1.0cm} | C{1.0cm}C{1.6cm}C{1.8cm}C{1.2cm} }
    \ChangeRT{1pt}
        & \textbf{\small{AUC}} & \textbf{\small{Recall@50}} & \textbf{\small{Recall@100}} & \textbf{Time} (h) \\
    \hline
       $\gamma=0$  & 73.70 & 8.11 & 12.73 & 38.5  \\
       $\gamma=10$ & \textbf{73.94} & \textbf{8.40} & {13.09} & 26.2  \\
       $\gamma=20$ & 73.74 & 8.32 & \textbf{13.17} & 19.4  \\
       $\gamma=30$ & 73.69 & 7.99 & 12.37 & 15.4  \\
    \ChangeRT{1pt}
    \end{tabular}

\end{table}

\begin{figure}[t]
\centering
\includegraphics[width=0.97\columnwidth]{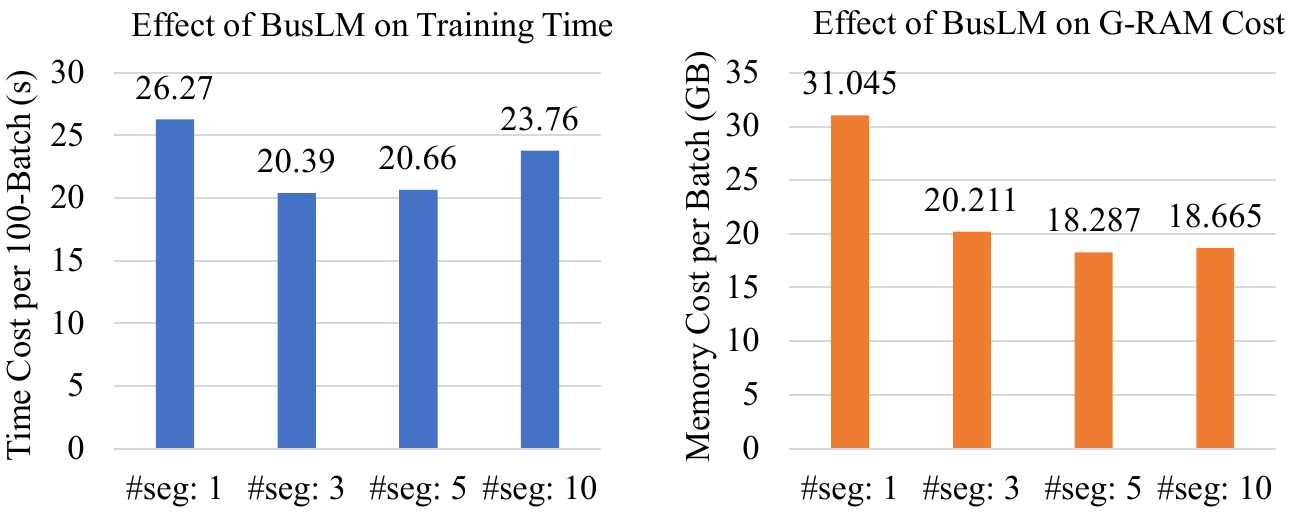}
\caption{BusLM's effect on training speed and GPU RAM usage, with \#seg increased from 1 to 10.}
\label{fig:exp-be}
\end{figure}

$\bullet$ \textbf{More on Cache}. 
The effect of cache-accelerated news encoding is tested with different values of $\gamma$ with Table \ref{tab:exp-cache} (the ``expiration step'' defined in Section \ref{sec:cache}). With the increment of $\gamma$, the training speed is accelerated tremendously. We choose ``$\gamma=20$'' as our default setting for the trade-off of training efficiency and quality. Besides, it is more interesting to see that the cache-accelerated news encoding also contributes to the recommendation quality. This is probably because the cache-accelerated news encoding helps to put more emphasis on the long-tailed news articles, which could suffer from insufficient training due to the dominance of the most popular news. In other words, most of the training opportunities would be taken by a small number of news articles (if cached is disabled), given that a large portion of the news clicks are resulted from the hottest minority (as reflected by Table \ref{tab:news-long}). 

$\bullet$ \textbf{More on BusLM}.
Shown as Figure \ref{fig:exp-be}, the training speed is improved and GPU RAM consumption is reduced thanks to BusLM (``\#seg:1'' means the input news remains one single sequence). However, it is also found that the training speed will no longer increase when \#seg is beyond 5. This is because the over-partition of the news will result in too many bus elements (discussed in Section \ref{sec:bus}), which magnifies the cost of information exchange across the segments. Specifically, it will slow down the layer-wise transformer encoding in Eq. \ref{eq:bus}. By default, the \#seg is set to 3 for the best performance. 



\section{Conclusion}
In this paper, we propose a novel framework, SpeedyFeed, for the efficient training of PLMs-based news recommender. SpeedyFeed enjoys three technical advantages: high reusability, high data efficiency, and economic news encoding complexity, which jointly lead to a huge speedup of the training workflow. The proposed framework is verified in Microsoft News, where significant improvements are achieved in comprehensive evaluations. The proposed framework is made public-available so as to facilitate the development in related areas. In the future, we'll proactively extend this framework to support more real-world applications, such as commodity and advertisement recommendation.

\appendices


\section{Bus Language Modeling}
\label{apx:blm}
The news article is split into $K$ segments which are interconnected by bus technology. For the $i$-th layer of segment $j$, the first tokens (i.e., [CLS]) is chosen to form the bus for information exchange, then the bus is spliced into each segment as the input of transformer layer:
\begin{equation}\label{eq:bus1}
    \mathbf{Bus}^i = \{\mathbf{H}^i_j[0]\}^K_{j=1},
\end{equation} 
\begin{equation}\label{eq:bus2}
    \mathbf{H}^{i+1}_j = \mathrm{Transformer}^{i}(\big[\mathbf{H}^i_j, ~\mathbf{Bus}^i\big]).
\end{equation}
In particular, for the self-attention layer of transformer, the bus is applied in key and value, and the query still only adopt the original embedding sequence of segments:
\begin{equation}
    \mathbf{Q}^{i}_j = \mathbf{H}^i_j\mathbf{W}^i_Q,
\end{equation}
\begin{equation}
    \mathbf{K}^{i}_j = (\big[\mathbf{H}^i_j, ~\mathbf{Bus}^i\big])\mathbf{W}^i_K,
\end{equation}
\begin{equation}
    \mathbf{V}^{i}_j = (\big[\mathbf{H}^i_j, ~\mathbf{Bus}^i\big])\mathbf{W}^i_V.
\end{equation}
In this way, $\mathbf{H}^{i+1}_j$ has the same shape as $\mathbf{H}^{i}_j$. We repeat the Eqns~\ref{eq:bus1} and~\ref{eq:bus2} for each layer.

After all transformer layers, we aggregate all of the hidden states in the last layer (i.e., $\mathbf{H}_{*}^{-1}$) as news embeddings by two additional attention layers.
Specifically, the first attention layer is proposed to  learn more informative
representations of segments. The attention weight $\mathbf{\alpha}_{j,n}$ of the $n$-th tokens in $j$-th segment is computed as:
\begin{equation}
    \mathbf{\alpha}_{j,n} = \mathbf{q}^T_1 tanh(\mathbf{W}_1 \mathbf{H}^{-1}_{j,n} + \mathbf{b}_1)
\end{equation}
\begin{equation}
    \mathbf{\alpha}_{j,n} = \frac{\exp(\mathbf{\alpha}_{j,n})}{\sum_{n=1}^L \exp(\mathbf{\alpha}_{j,n})}
\end{equation}
where $\mathbf{W}_1$ and $\mathbf{b}_1$ are projection parameters, and $\mathbf{q}^T_1$ is the query vector. The representation of
segment is the weighted summation of the contextual tokens representations, formulated as:
\begin{equation}
    \mathbf{v}_j = \sum_{n=1}^L\mathbf{\alpha}_{j,n} \mathbf{H}^{-1}_{j,n}
\end{equation}
The second attention layer is to aggregate the segment embedding $\mathbf{v}_j$. Similarly, denote the attention weight of the $j$-th
segment as $\mathbf{\alpha}_{j}$, which is calculated by:
\begin{equation}
    \mathbf{\alpha}_{j} = \mathbf{q}^T_2 tanh(\mathbf{W}_2 \mathbf{v}_{j} + \
    \mathbf{b}_2)
\end{equation}
\begin{equation}
    \mathbf{\alpha}_{j} = \frac{\exp(\mathbf{\alpha}_{j})}{\sum_{j=1}^K \exp(\mathbf{\alpha}_{j})}
\end{equation}
where $\mathbf{q}^T_2$, $\mathbf{W}_2$, and $\mathbf{b}_2$ are learnable parameters. The final representation of a news
article is the summation of the segment representations
weighted by their attention weights: 
\begin{equation}
    \mathbf{e} = \sum_{j=1}^K\mathbf{\alpha}_{j} \mathbf{v}_{j}
\end{equation}

\section{Details of Dataset}\label{apd:dataset}
$\bullet$ \textbf{News}. Each news article is composed of its title, abstract and body. The average text length is 659.64, which is even longer than the default maximum length (512) of ordinary PLMs. It is also challenging to load a sufficient amount of training instances into a mini-batch given such long texts. However, knowing that the title, abstract and the 1st paragraph in the body are the most informative parts for the majority of Microsoft news, we'll take a text segment from each of them, whose length is no more than 32. Finally, the overall text length is confined within 96 for the trade-off of quality and feasibility.

$\bullet$ \textbf{User}. The users are characterized by their historical interaction with the platform. As a result, each user is associated with one record, containing all of the user's impressions ordered by time: 
\begin{center}
``{User-ID} \# {Impression-0} ... {Impression-N}''.
\end{center}
Each impression refers to user's news browsing behaviors within one interaction, which is in the format of: 
\begin{center}
``{Impression-ID} \# {Time} \# {Clicked-list} \# {Impressed-list}''.
\end{center}
The clicked list refers to all of the news articles clicked by the user in the impression, the impressed list includes the news articles shown to user but not clicked. The user's activeness follows long-tail distribution: each user has 15.27 news clicks on average; however, 
3.26\% users have more than 100 news clicks, and 
1.44\% users have more than 150 news clicks. In our experiments, the user history is truncated to 100 uniformly.

\section{Details of Training Configurations}\label{apd:train}
All the training jobs are performed on an Azure\footnote{https://azure.microsoft.com/en-us/services/machine-learning/} machine, with 4$\times$Nvidia-V100-32G GPUs, 40$\times$Intel(R) Xeon(R) Platinum 8168 CPU @ 2.70GHz processors, run on Ubuntu 16.04.6. The models are implemented with PyTorch 1.7.0. 
We optimize the parameters with the Adam optimizer. The learning rate is 8e-6 for pretrained model and 1e-4 for other layers (e.g., user encoder).
The negative sampling ratio is 1. The max length of user click history is 100. The default max length of news article is 96 and we split the text into three segments according to the structure of title, abstract and body.
The default pretrained model is the complete UniLM. We use 2 buckets and push the train data in buckets to the mini-batch queue once a basket is filled with 39800 tokens. For content refinement, the $k_1$ of BM25 is 2 and we reserve the words of top-32 BM25 scores for each segment. For cache management policy, the hyper-parameter $\beta$ is 2e-3, and the default expiration-step $\gamma$ of cache is 20.




\bibliographystyle{IEEEtran}
\bibliography{ref}

\begin{thebibliography}{10}
\providecommand{\url}[1]{#1}
\csname url@samestyle\endcsname
\providecommand{\newblock}{\relax}
\providecommand{\bibinfo}[2]{#2}
\providecommand{\BIBentrySTDinterwordspacing}{\spaceskip=0pt\relax}
\providecommand{\BIBentryALTinterwordstretchfactor}{4}
\providecommand{\BIBentryALTinterwordspacing}{\spaceskip=\fontdimen2\font plus
\BIBentryALTinterwordstretchfactor\fontdimen3\font minus
  \fontdimen4\font\relax}
\providecommand{\BIBforeignlanguage}[2]{{%
\expandafter\ifx\csname l@#1\endcsname\relax
\typeout{** WARNING: IEEEtran.bst: No hyphenation pattern has been}%
\typeout{** loaded for the language `#1'. Using the pattern for}%
\typeout{** the default language instead.}%
\else
\language=\csname l@#1\endcsname
\fi
#2}}
\providecommand{\BIBdecl}{\relax}
\BIBdecl

\bibitem{wu2020mind}
F.~Wu, Y.~Qiao, J.-H. Chen, C.~Wu, T.~Qi, J.~Lian, D.~Liu, X.~Xie, J.~Gao,
  W.~Wu \emph{et~al.}, ``Mind: A large-scale dataset for news recommendation,''
  in \emph{Proceedings of the 58th Annual Meeting of the Association for
  Computational Linguistics}, 2020, pp. 3597--3606.

\bibitem{okura2017embedding}
S.~Okura, Y.~Tagami, S.~Ono, and A.~Tajima, ``Embedding-based news
  recommendation for millions of users,'' in \emph{Proceedings of the 23rd ACM
  SIGKDD International Conference on Knowledge Discovery and Data Mining},
  2017, pp. 1933--1942.

\bibitem{li2010contextual}
L.~Li, W.~Chu, J.~Langford, and R.~E. Schapire, ``A contextual-bandit approach
  to personalized news article recommendation,'' in \emph{Proceedings of the
  19th international conference on World wide web}, 2010, pp. 661--670.

\bibitem{devlin2018bert}
J.~Devlin, M.-W. Chang, K.~Lee, and K.~Toutanova, ``Bert: Pre-training of deep
  bidirectional transformers for language understanding,'' \emph{arXiv preprint
  arXiv:1810.04805}, 2018.

\bibitem{liu2019roberta}
Y.~Liu, M.~Ott, N.~Goyal, J.~Du, M.~Joshi, D.~Chen, O.~Levy, M.~Lewis,
  L.~Zettlemoyer, and V.~Stoyanov, ``Roberta: A robustly optimized bert
  pretraining approach,'' \emph{arXiv preprint arXiv:1907.11692}, 2019.

\bibitem{das2007google}
A.~S. Das, M.~Datar, A.~Garg, and S.~Rajaram, ``Google news personalization:
  scalable online collaborative filtering,'' in \emph{Proceedings of the 16th
  international conference on World Wide Web}, 2007, pp. 271--280.

\bibitem{covington2016deep}
P.~Covington, J.~Adams, and E.~Sargin, ``Deep neural networks for youtube
  recommendations,'' in \emph{Proceedings of the 10th ACM conference on
  recommender systems}, 2016, pp. 191--198.

\bibitem{hidasi2015session}
B.~Hidasi, A.~Karatzoglou, L.~Baltrunas, and D.~Tikk, ``Session-based
  recommendations with recurrent neural networks,'' \emph{arXiv preprint
  arXiv:1511.06939}, 2015.

\bibitem{zhou2018deep}
G.~Zhou, X.~Zhu, C.~Song, Y.~Fan, H.~Zhu, X.~Ma, Y.~Yan, J.~Jin, H.~Li, and
  K.~Gai, ``Deep interest network for click-through rate prediction,'' in
  \emph{Proceedings of the 24th ACM SIGKDD International Conference on
  Knowledge Discovery \& Data Mining}, 2018, pp. 1059--1068.

\bibitem{zhou2019deep}
G.~Zhou, N.~Mou, Y.~Fan, Q.~Pi, W.~Bian, C.~Zhou, X.~Zhu, and K.~Gai, ``Deep
  interest evolution network for click-through rate prediction,'' in
  \emph{Proceedings of the AAAI conference on artificial intelligence},
  vol.~33, no.~01, 2019, pp. 5941--5948.

\bibitem{chen2018sequential}
X.~Chen, H.~Xu, Y.~Zhang, J.~Tang, Y.~Cao, Z.~Qin, and H.~Zha, ``Sequential
  recommendation with user memory networks,'' in \emph{Proceedings of the
  eleventh ACM international conference on web search and data mining}, 2018,
  pp. 108--116.

\bibitem{wang2018dkn}
H.~Wang, F.~Zhang, X.~Xie, and M.~Guo, ``Dkn: Deep knowledge-aware network for
  news recommendation,'' in \emph{Proceedings of the 2018 world wide web
  conference}, 2018, pp. 1835--1844.

\bibitem{an2019neural}
M.~An, F.~Wu, C.~Wu, K.~Zhang, Z.~Liu, and X.~Xie, ``Neural news recommendation
  with long-and short-term user representations,'' in \emph{Proceedings of the
  57th Annual Meeting of the Association for Computational Linguistics}, 2019,
  pp. 336--345.

\bibitem{yang2016hierarchical}
Z.~Yang, D.~Yang, C.~Dyer, X.~He, A.~Smola, and E.~Hovy, ``Hierarchical
  attention networks for document classification,'' in \emph{Proceedings of the
  2016 conference of the North American chapter of the association for
  computational linguistics: human language technologies}, 2016, pp.
  1480--1489.

\bibitem{wu2019neural}
C.~Wu, F.~Wu, M.~An, J.~Huang, Y.~Huang, and X.~Xie, ``Neural news
  recommendation with attentive multi-view learning,'' \emph{arXiv preprint
  arXiv:1907.05576}, 2019.

\bibitem{zhou2018multi}
X.~Zhou, L.~Li, D.~Dong, Y.~Liu, Y.~Chen, W.~X. Zhao, D.~Yu, and H.~Wu,
  ``Multi-turn response selection for chatbots with deep attention matching
  network,'' in \emph{Proceedings of the 56th Annual Meeting of the Association
  for Computational Linguistics (Volume 1: Long Papers)}, 2018, pp. 1118--1127.

\bibitem{mikolov2013distributed}
T.~Mikolov, I.~Sutskever, K.~Chen, G.~Corrado, and J.~Dean, ``Distributed
  representations of words and phrases and their compositionality,''
  \emph{arXiv preprint arXiv:1310.4546}, 2013.

\bibitem{pennington2014glove}
J.~Pennington, R.~Socher, and C.~D. Manning, ``Glove: Global vectors for word
  representation,'' in \emph{Proceedings of the 2014 conference on empirical
  methods in natural language processing (EMNLP)}, 2014, pp. 1532--1543.

\bibitem{peters2018deep}
M.~E. Peters, M.~Neumann, M.~Iyyer, M.~Gardner, C.~Clark, K.~Lee, and
  L.~Zettlemoyer, ``Deep contextualized word representations,'' \emph{arXiv
  preprint arXiv:1802.05365}, 2018.

\bibitem{radford2018improving}
A.~Radford, K.~Narasimhan, T.~Salimans, and I.~Sutskever, ``Improving language
  understanding by generative pre-training,'' \emph{Technical report, OpenAI},
  2018.

\bibitem{bao2020unilmv2}
H.~Bao, L.~Dong, F.~Wei, W.~Wang, N.~Yang, X.~Liu, Y.~Wang, J.~Gao, S.~Piao,
  M.~Zhou \emph{et~al.}, ``Unilmv2: Pseudo-masked language models for unified
  language model pre-training,'' in \emph{International Conference on Machine
  Learning}.\hskip 1em plus 0.5em minus 0.4em\relax PMLR, 2020, pp. 642--652.

\bibitem{brown2020language}
T.~B. Brown, B.~Mann, N.~Ryder, M.~Subbiah, J.~Kaplan, P.~Dhariwal,
  A.~Neelakantan, P.~Shyam, G.~Sastry, A.~Askell \emph{et~al.}, ``Language
  models are few-shot learners,'' \emph{arXiv preprint arXiv:2005.14165}, 2020.

\bibitem{chang2020pre}
W.-C. Chang, F.~X. Yu, Y.-W. Chang, Y.~Yang, and S.~Kumar, ``Pre-training tasks
  for embedding-based large-scale retrieval,'' \emph{arXiv preprint
  arXiv:2002.03932}, 2020.

\bibitem{humeau2019poly}
S.~Humeau, K.~Shuster, M.-A. Lachaux, and J.~Weston, ``Poly-encoders:
  Transformer architectures and pre-training strategies for fast and accurate
  multi-sentence scoring,'' \emph{arXiv preprint arXiv:1905.01969}, 2019.

\bibitem{khattab2020colbert}
O.~Khattab and M.~Zaharia, ``Colbert: Efficient and effective passage search
  via contextualized late interaction over bert,'' in \emph{Proceedings of the
  43rd International ACM SIGIR Conference on Research and Development in
  Information Retrieval}, 2020, pp. 39--48.

\bibitem{guu2020realm}
K.~Guu, K.~Lee, Z.~Tung, P.~Pasupat, and M.-W. Chang, ``Realm:
  Retrieval-augmented language model pre-training,'' \emph{arXiv preprint
  arXiv:2002.08909}, 2020.

\bibitem{lu2020twinbert}
W.~Lu, J.~Jiao, and R.~Zhang, ``Twinbert: Distilling knowledge to
  twin-structured compressed bert models for large-scale retrieval,'' in
  \emph{Proceedings of the 29th ACM International Conference on Information \&
  Knowledge Management}, 2020, pp. 2645--2652.

\bibitem{msreco}
\BIBentryALTinterwordspacing
\emph{Microsoft Recommenders}, 2020. [Online]. Available:
  \url{https://github.com/microsoft/recommenders/}
\BIBentrySTDinterwordspacing

\bibitem{vaswani2017attention}
A.~Vaswani, N.~Shazeer, N.~Parmar, J.~Uszkoreit, L.~Jones, A.~N. Gomez,
  {\L}.~Kaiser, and I.~Polosukhin, ``Attention is all you need,'' in
  \emph{Advances in neural information processing systems}, 2017, pp.
  5998--6008.

\bibitem{tay2020efficient}
Y.~Tay, M.~Dehghani, D.~Bahri, and D.~Metzler, ``Efficient transformers: A
  survey,'' \emph{arXiv preprint arXiv:2009.06732}, 2020.

\bibitem{wang2020minilm}
W.~Wang, F.~Wei, L.~Dong, H.~Bao, N.~Yang, and M.~Zhou, ``Minilm: Deep
  self-attention distillation for task-agnostic compression of pre-trained
  transformers,'' \emph{arXiv preprint arXiv:2002.10957}, 2020.

\bibitem{wu2019npa}
C.~Wu, F.~Wu, M.~An, J.~Huang, Y.~Huang, and X.~Xie, ``Npa: neural news
  recommendation with personalized attention,'' in \emph{Proceedings of the
  25th ACM SIGKDD International Conference on Knowledge Discovery \& Data
  Mining}, 2019, pp. 2576--2584.

\bibitem{wu2019nrms}
C.~Wu, F.~Wu, S.~Ge, T.~Qi, Y.~Huang, and X.~Xie, ``Neural news recommendation
  with multi-head self-attention,'' in \emph{Proceedings of the 2019 Conference
  on Empirical Methods in Natural Language Processing and the 9th International
  Joint Conference on Natural Language Processing (EMNLP-IJCNLP)}, Hong Kong,
  China, Nov. 2019, pp. 6389--6394.

\bibitem{malkov2018efficient}
Y.~A. Malkov and D.~A. Yashunin, ``Efficient and robust approximate nearest
  neighbor search using hierarchical navigable small world graphs,'' \emph{IEEE
  transactions on pattern analysis and machine intelligence}, vol.~42, no.~4,
  pp. 824--836, 2018.

\end{thebibliography}





\end{document}